\documentclass[aps,prl,reprint,groupedaddress]{revtex4-1}
\usepackage{amssymb}
\usepackage{amsmath}
\usepackage{array}
\usepackage{multirow,bigdelim}
\usepackage{enumitem}

\begin{document}

\title{Partition GHZ SLOCC class of three qubits into ten families under LU}
\author{Dafa Li}

\begin{abstract}
In [Science 340:1205, (2013)], via entanglement polytopes Michael Walter et
al. obtained a finite yet systematic classification of multi-particle
entanglement. It is well known that under SLOCC, pure states of three (four)
qubits are partitioned into six (nine) families. In this paper,we present a
LU invariant and an entanglement measures for the GHZ SLOCC class of three
qubits, and partition states of the GHZ SLOCC class of three qubits into ten
families and each family into two subfamilies under LU. We give a necessary
and sufficient condition for the uniqueness of the generalized Schmidt
decomposition for the GHZ SLOCC class.
\end{abstract}


\affiliation{
Department of Mathematical Sciences, Tsinghua University,
Beijing, 100084, China}


\maketitle

\section{Introduction}

Quantum entanglement is considered as a key quantum mechanical resource in
quantum information and computation such as quantum teleportation, quantum
cryptography, quantum metrology, and quantum key distribution \cite{Nielsen}%
. To understand entanglement, lots of efforts have contributed to study the
convertibility of two states under local unitary operators (LU), local
operations and classical communication (LOCC), and Stochastic LOCC (SLOCC).

Two pure $n$-qubit states $|\psi ^{\prime }\rangle $ and $|\psi \rangle $
are LU (SLOCC) equivalent if the two states satisfy the following equation,

\begin{equation}
|\psi ^{\prime }\rangle =A_{1}\otimes A_{2}\otimes \cdots \otimes A_{n}|\psi
\rangle ,
\end{equation}%
where $2$ by $2$ matrices $A_{i}$ are unitary (invertible).

Under SLOCC, pure states of three qubits were distinguished into six
equivalence classes GHZ, W, A-BC, B-AC, C-AB, and A-B-C \cite{Dur}, and pure
states of four qubits were partitioned into nine families or more \cite%
{Verstraete, dli-qip20}. Classification of multipartite entangled states by
multidimensional determinants were investigated \cite{Miyake}. It is known
that if two states are LU equivalent, then they have the same amount of
entanglement and can do the same tasks in quantum information theory \cite%
{Dur, Verstraete, Kraus-prl, Kraus-PRA}.

Lots of efforts have devoted to studying the characterization, the
quantification, and the classification of the entanglement via Schmidt
decomposition \cite{Acin00, Vicente, Sudbery, Leifer, Sabin, Grassl,
Carteret, Soeda, Tamaryan, Partovi, Tyson, Liu-12, Jun-li, Dli-qip-18,
dli-jpa, Kumari, Adhikari}. Ac\'{\i}n et al. proposed the generalized
Schmidt decomposition\ of three qubits \cite{Acin00}. Carteret et al.
discuss the Schmidt decomposition for the multipartite system \cite{Carteret}%
. Kraus proved that two states are LU equivalent if and only if they have
the same standard forms \cite{Kraus-prl, Kraus-PRA}. Vicente et al. derived
a new decomposition for pure states of three qubits, which is characterized
by five parameters (up to local unitary operations) \cite{Vicente}. Liu et
al. proposed a practical entanglement classification scheme for pure states
of general multipartite for arbitrary dimensions under LU \cite{Liu-12}. Li
and Qiao proposed a practical method for finding the canonical forms for
pure and mixed states of arbitrary dimensional multipartite systems under LU
\cite{Jun-li}. Via the generalized Schmidt decomposition\ of three qubits,
Kumari and Adhikari partitioned positive states (i.e. the states with the
phase factor $\theta =0$) of the GHZ SLOCC\ class into four subclasses and
proposed the witness operator for the classification \cite{Kumari}.

The canonical form and the entanglement measure and classification for three
qubits have been widely studied topics \cite{Acin00, Enriquez, Tajima,
Liu-15, Dli-qip-18, dli-jpa, Salinas, Vicente, Kumari, Adhikari}. The SLOCC\
entanglement classification of three and four qubits has many applications.
For example, a one-to-one correspondence between the SLOCC\ entanglement
classification of three and four qubits and the classification of the
extremal black holes was discussed \cite{black-hole}. LU classification of
black holes corresponding to GHZ SLOCC\ class is studied in \cite{bh-ghz}.%
\newline
\ \ \ \ \ In this paper, we present a LU\ invariant for the GHZ\ SLOCC\
class of three qubits, and partition the GHZ\ SLOCC\ class of three qubits
into ten families and each family into two subfamilies under LU. Thus, the
infinite LU equivalence classes of the GHZ SLOCC\ classes are partitioned
into 20 subfamilies.

\section{Parameters $\protect\varrho $ and $\protect\iota $ for the GHZ
SLOCC\ class}

By means of LU transformations, any pure state of three qubits can be
written as

\begin{eqnarray}
|\psi \rangle &=&\lambda _{0}|000\rangle +\lambda _{1}e^{i\varphi
}|100\rangle  \notag \\
&&+\lambda _{2}|101\rangle +\lambda _{3}|110\rangle +\lambda _{4}|111\rangle
,  \label{lu-ghz-s-1}
\end{eqnarray}%
where $\lambda _{i}\geq 0$, $\sum_{i=0}^{4}\lambda _{i}^{2}=1$, $0\leq
\varphi \leq \pi $, $\varphi $ is referred to as the phase of $|\psi \rangle
$ \cite{Acin00}. In this paper, $\varphi $ is limited to $[0,2\pi )$. Eq. (%
\ref{lu-ghz-s-1}) is referred to as Ac\'{\i}n et al.'s Schmidt Decomposition
(ASD) of the state $|\psi \rangle $.

For simplicity, $|\psi \rangle $ is written as
\begin{equation}
|\psi \rangle =(\lambda _{0},\lambda _{1}e^{i\varphi },\lambda _{2},\lambda
_{3},\lambda _{4}),  \label{lu-ghz-2}
\end{equation}%
which is the set of the coefficients of the five LBPS (local bases product
states). The set of \ the coefficients in Eq. (\ref{lu-ghz-2}) is called Ac%
\'{\i}n et al.'s Schmidt coefficients (ASC)\ of $\ |\psi \rangle $.

A state is referred as to an $i$-LBPS state if the state has just $i$
non-vanishing Schmidt coefficients.

Motivation 1. In \cite{Acin00}, pure states of three qubits were partitioned
into five types. Specially, states of the GHZ SLOCC class were partitioned
into Types 2b, 3b, 4b, 4c, and 5. The Schmidt Decomposition for three qubits
and the LU classification of three qubits have had a significant impact on
QIC.

It is well known that in several aspects the GHZ state $\frac{1}{\sqrt{2}}%
(|000\rangle +|111\rangle )$ can be regarded as the maximally entangled
state of three qubits \cite{Dur}. Recently, Kumari and Adhikari divided
positive states (i.e. the states with the vanishing phases) of the GHZ SLOCC
class into four subclasses 1, 2, 3, 4 and compared the maximal teleportation
fidelities, the entanglement, and the tangle in the four subclasses. Of
course, it is fundamental to partition the GHZ SLOCC\ class completely under
LU.

Motivation 2. We want to find a criterion to determine what states have the
unique Schmidt decomposition or not. For the states which have the unique
Schmidt decomposition, one can see that subjected to local random unitary
noise, their ASDs do not change.

Motivation 3. It is interesting to study the classification for back holes
via the LU entanglement classification of the GHZ SLOCC\ class \cite{bh-ghz}%
. Lot of research had been done on the relation between the SLOCC\
entanglement classification of three and four qubits and the classification
of the extremal black holes \cite{black-hole}.

Kallosh and Linde investigated the black holes with\ 4 non-vanishing integer
charges $q_{0}$, $p^{1}$, $p^{2}$, $p^{3}$, which correspond to the
following states \cite{Linde}.

\begin{equation}
-p^{1}|001\rangle -p^{2}|010\rangle -p^{3}|100\rangle +q_{0}|111\rangle .
\label{ghz-1}
\end{equation}

The states in Eq. (\ref{ghz-1}) belong to the GHZ SLOCC class. It is
potential to establish a relation between the classification of the black
holes with 4 non-vanishing charges $q_{0}$, $p^{1}$, $p^{2}$, $p^{3}$ and
the LU classification of the GHZ SLOCC\ class \cite{bh-ghz}.

We next discuss LU\ classification of the GHZ SLOCC equivalence\ class via
ASD. $|\psi \rangle $ in Eq. (\ref{lu-ghz-s-1}) belongs to the GHZ\ SLOCC\
class if and only if $\lambda _{0}\lambda _{4}\neq 0$ \cite{Dli-qip-18}. \

Let
\begin{equation}
\gamma =\lambda _{1}\lambda _{4}e^{i\varphi }-\lambda _{2}\lambda _{3}.
\label{gama-1}
\end{equation}%
The parameters $\varrho $ and $\iota $ were defined for the states with $%
\gamma \neq 0$ of the GHZ SLOCC\ class \cite{dli-jpa}. Here, we define the
parameters $\varrho $ and $\iota $ for the whole GHZ SLOCC\ class, then show
that $|\ln \varrho |$ is a LU invariant for the whole GHZ SLOCC\ class below.

For $|\psi \rangle $ in Eq. (\ref{lu-ghz-s-1}), when $\lambda _{0}\lambda
_{4}\neq 0$, we define
\begin{eqnarray}
\varrho (|\psi \rangle ) &=&\sqrt{J_{4}+J_{1}}/\sqrt{(\lambda
_{2}^{2}+\lambda _{4}^{2})(\lambda _{3}^{2}+\lambda _{4}^{2})},  \label{fact}
\\
\iota (|\psi \rangle ) &=&(\lambda _{2}\lambda _{3}+\gamma ^{\ast }/\varrho
^{2})/\lambda _{4},  \label{factor}
\end{eqnarray}%
\ where $\gamma ^{\ast }$ is the complex conjugate of $\gamma $, $%
J_{1}=|\gamma |^{2}$, and $J_{4}=(\lambda _{0}\lambda _{4})^{2}$. $J_{1}$
and $J_{4}$\ are LU invariants \cite{Acin00}. Clearly, $\varrho (|\psi
\rangle )>0$.

When it is clear from the context, we write $\varrho $ and $\iota $ for $%
\varrho (|\psi \rangle )$ and $\iota (|\psi \rangle )$, respectively. $%
\varrho $ and $\iota $ are used to describe the LU equivalence of two ASD
states and partition ASD states of the GHZ SLOCC\ class under LU below.

Let us consider the state

\begin{equation}
|\psi _{\varrho ,\iota }\rangle =((1/\varrho )\lambda _{0},\varrho \iota
,\varrho \lambda _{2},\varrho \lambda _{3},\varrho \lambda _{4}).
\label{eq-state-1}
\end{equation}

We say that $|\psi _{\varrho ,\iota }\rangle $ is obtained by applying $%
\varrho -\iota $ transformation to $|\psi \rangle $. Note that the phase of
the complex number $\iota $ is just the phase of $|\psi _{\varrho ,\iota
}\rangle $. Let $\varrho ^{\prime }=\varrho (|\psi _{\varrho ,\iota }\rangle
)$ and $\iota ^{\prime }=\iota (|\psi _{\varrho ,\iota }\rangle )$. Then, a
calculation yields

\begin{equation}
\varrho ^{\prime }=1/\varrho .  \label{prop-1}
\end{equation}%
That is,
\begin{equation}
\varrho ^{\prime }\varrho =1.  \label{invar}
\end{equation}

Via Eqs. (\ref{factor}, \ref{prop-1}),
\begin{equation}
\iota ^{\prime }=\varrho \lambda _{1}e^{i\varphi }.  \label{iota-3}
\end{equation}

We next show that $|\psi \rangle $ can also be obtained by applying $\varrho
^{\prime }-\iota ^{\prime }$ transformation to $|\psi _{\varrho ,\iota
}\rangle $ in Eq. (\ref{eq-state-1}). From Eqs. (\ref{eq-state-1}, \ref%
{prop-1}, \ref{iota-3}), a calculation yields that

\begin{eqnarray}
&&((1/\varrho ^{\prime })((1/\varrho )\lambda _{0}),\varrho ^{\prime }\iota
^{\prime },\varrho ^{\prime }(\varrho \lambda _{2}),\varrho ^{\prime
}(\varrho \lambda _{3}),\varrho ^{\prime }(\varrho \lambda _{4}))  \notag \\
&=&(\lambda _{0},\lambda _{1}e^{i\varphi },\lambda _{2},\lambda _{3},\lambda
_{4})=|\psi \rangle .  \label{eq-4}
\end{eqnarray}

Therefore, if $|\psi \rangle $ can be $\varrho -\iota $ transformed into $%
|\psi _{\varrho ,\iota }\rangle $, then $|\psi _{\varrho ,\iota }\rangle $
can also be $\varrho ^{\prime }-\iota ^{\prime }$ transformed into $|\psi
\rangle $.

\section{LU Partition of the GHZ\ SLOCC\ class via $\protect\varrho $, $%
\protect\iota $, and $\protect\gamma $}

It is known that $\lambda _{0}\lambda _{4}\neq 0$ for the GHZ SLOCC\ class.
Here, the states with non-negative (real and complex) coefficients are
called positive (real and complex) states.

\subsection{LU classification of positive states with $\protect\gamma =0$}

\subsubsection{Calculating $\protect\varrho $, $\protect\iota $, and $|%
\protect\psi _{\protect\varrho ,\protect\iota }\rangle $}

When $\gamma =0$, $|\psi \rangle $ in Eq. (\ref{lu-ghz-2}) can be written as
\begin{equation}
|\psi \rangle =(\lambda _{0},\lambda _{1},\lambda _{2},\lambda _{3},\lambda
_{4}),  \label{p-1}
\end{equation}%
where $\lambda _{1}\lambda _{4}=\lambda _{2}\lambda _{3}$ \cite{dli-jpa}.
For $|\psi \rangle $ in Eq. (\ref{p-1}), a calculation yields that
\begin{eqnarray}
\varrho &=&\lambda _{0}/\sqrt{1-\lambda _{0}^{2}},  \label{sigma-1} \\
\iota &=&\lambda _{2}\lambda _{3}/\lambda _{4}=\lambda _{1}\lambda
_{4}/\lambda _{4}=\lambda _{1},  \label{uota-1} \\
|\psi _{\varrho ,\iota }\rangle &=&((1/\varrho )\lambda _{0},\varrho \lambda
_{1},\varrho \lambda _{2},\varrho \lambda _{3},\varrho \lambda _{4}).
\label{prop-2}
\end{eqnarray}%
Specially, when $\varrho =1$, from Eqs. (\ref{sigma-1}, \ref{prop-2}), we
obtain
\begin{eqnarray}
\lambda _{0} &=&\frac{1}{\sqrt{2}}, \\
|\psi _{\varrho ,\iota }\rangle &=&|\psi \rangle .\
\end{eqnarray}

\textit{Result 1.} In light of Proposition 2 in \cite{dli-jpa}, one can know
that $|\psi ^{\prime }\rangle $ is LU equivalent to $|\psi \rangle $ with $%
\gamma =0$ if and only if $\ |\psi ^{\prime }\rangle =|\psi _{\varrho ,\iota
}\rangle $ in Eq. (\ref{prop-2}).

From Result 1, we have the following corollary 1.1.

\textit{Corollary 1.1}. If $|\psi ^{\prime }\rangle $ is LU equivalent to $%
|\psi \rangle $ with $\gamma =0$, then $|\ln \varrho |=|\ln \varrho ^{\prime
}|$, and $|\psi ^{\prime }\rangle $ and $|\psi \rangle $ both are positive
and have the same kinds of LBPS.

\subsubsection{LU classification of positive states with $\protect\gamma =0$}

The states with $\gamma =0$ are partitioned into four positive families $%
P_{i}$, $i=1,\cdots ,4$. Ref. Table I.

We next argue that $P_{i}$, $i=1,\cdots ,4$, are LU inequivalent.

In light of Result 1 and via Eqs. (\ref{p-1}, \ref{prop-2}), one can see
that $\lambda _{i}$ and $\varrho \lambda _{i}$, $i=1,2,3$, both vanish or
neither does. It guarantees that $P_{i}$, $i=1,\cdots ,4$, are LU
inequivalent. For example, $\lambda _{1}\neq 0$ for $P_{1}$ while $\lambda
_{1}=0$ for $P_{i}$, $i=2,3,4$. Therefore, $P_{1}$ is LU inequivalent to $%
P_{i}$, $i=2,3,4$.

Again, $P_{i}$ is divided into two subfamilies $P_{i}^{\prime }$ (states
with $\varrho =1$) and $P_{i}^{\prime \prime }$ (states with $\varrho \neq 1$%
). Ref. Table 1. Corollary 1.1 implies that $P_{i}^{\prime }$ and $%
P_{i}^{\prime \prime }$, $i=1,2,3,4$, are LU\ inequivalent. Clearly, each LU
class of $P_{i}^{\prime }$ is a singleton, and each LU class of $%
P_{i}^{\prime \prime }$ consists of only two states $|\psi \rangle $ and $%
|\psi _{\varrho ,\iota }\rangle $.\

Via Eq. (\ref{sigma-1}), a calculation yields that $\lambda _{0}=\frac{1}{%
\sqrt{2}}$ if and only if $\varrho =1$. In light of Result 1 and Corollary
1.1, we have the following Corollary 1.2.

\textit{Corollary 1.2}. ASD of a positive state with $\gamma =0$ is unique
if and only if $\varrho =1$. In other words, ASD of a positive state with $%
\gamma =0$ is unique if and only if $\lambda _{0}=\frac{1}{\sqrt{2}}$.

The contrapositive version of Corollary 1.2 leads to the following. ASD of a
positive state with $\gamma =0$ is not unique if and only if $\varrho \neq 1$
(in other words, $\lambda _{0}\neq \frac{1}{\sqrt{2}}$).

Table I. Positive families $P_{i}$, $i=1,\cdots ,4$, for which $\gamma =0$

$%
\begin{tabular}{|l|l|l|}
\hline
& $\varrho =1$ & $\varrho \neq 1$ \\ \hline
$P_{1}$; \{$(\lambda _{0},\lambda _{1},\lambda _{2},\lambda _{3},\lambda
_{4})$\}$\ddagger $ & $P_{1}^{\prime }$; $\lambda _{0}=\frac{1}{\sqrt{2}}$ &
$P_{1}^{\prime \prime }$ \\ \hline
$P_{2}$; \{$(\lambda _{0},0,0,\lambda _{3},\lambda _{4})$\} & $P_{2}^{\prime
}$; $\lambda _{0}=\frac{1}{\sqrt{2}}$ & $P_{2}^{\prime \prime }$ \\ \hline
$P_{3}$;\{$(\lambda _{0},0,\lambda _{2},0,\lambda _{4})$\} & $P_{3}^{\prime
} $; $\lambda _{0}=\frac{1}{\sqrt{2}}$ & $P_{3}^{\prime \prime }$ \\ \hline
$P_{4}$; \{$(\lambda _{0},0,0,0,\lambda _{4})$\} & $P_{4}^{\prime }=$\{GHZ\}
& $P_{4}^{\prime \prime }$ \\ \hline
\end{tabular}%
$

$\ddagger $ Each state of $P_{1}$ satisfies that $\lambda _{1}\lambda
_{4}=\lambda _{2}\lambda _{3}\neq 0$.

\subsection{LU classification of real states with $\protect\gamma \protect%
\lambda _{2}\protect\lambda _{3}\neq 0$}

Let
\begin{equation}
|\psi \rangle =(\lambda _{0},\delta \lambda _{1},\lambda _{2},\lambda
_{3},\lambda _{4}),  \label{real-st}
\end{equation}%
where $\delta =\pm 1$, $\gamma \neq 0$, $\lambda _{2}\lambda _{3}\neq 0$,
but $\lambda _{1}$ may vanish.

\subsubsection{Calculating $\protect\varrho $, $\protect\iota $, and $|%
\protect\psi _{\protect\varrho ,\protect\iota }\rangle $}

A calculation yields that
\begin{eqnarray}
\gamma &=&\delta \lambda _{1}\lambda _{4}-\lambda _{2}\lambda _{3},
\label{gam-1} \\
\varrho &=&\sqrt{J_{4}+J_{1}}/\sqrt{(\lambda _{2}^{2}+\lambda
_{4}^{2})(\lambda _{3}^{2}+\lambda _{4}^{2})}  \label{uniq-1} \\
\iota &=&(\lambda _{2}\lambda _{3}+\gamma /\varrho ^{2})/\lambda _{4},
\label{rel-2} \\
|\psi _{\varrho ,\iota }\rangle &=&((1/\varrho )\lambda _{0},\varrho \iota
,\varrho \lambda _{2},\varrho \lambda _{3},\varrho \lambda _{4}).
\label{rel-3}
\end{eqnarray}%
One can know that $\iota $ is real and $|\psi _{\varrho ,\iota }\rangle $ is
a real state.

When $\varrho =1$, from Eqs. (\ref{uniq-1}, \ref{rel-2}, \ref{rel-3}), a
calculation yields that
\begin{eqnarray}
\iota &=&\delta \lambda _{1},  \label{iot--} \\
\lambda _{0}^{2}+\lambda _{1}^{2} &=&\frac{1}{2}+\frac{\delta \lambda
_{1}\lambda _{2}\lambda _{3}}{\lambda _{4}},  \label{asd-unique-1} \\
|\psi _{\varrho ,\iota }\rangle &=&|\psi \rangle .  \label{unique-3}
\end{eqnarray}

\textit{Conclusion 1}. (i). Via Eq. (\ref{iot--}), we can conclude that for
a real 5-LBPS\ state with $\gamma \neq 0$, if $\varrho =1$, then $\iota \neq
0$ because $\lambda _{1}\neq 0$.

(ii). The contrapositive version of the above (i) leads to the following.
For a real 5-LBPS\ state with $\gamma \neq 0$, if $\iota =0$ then $\varrho
\neq 1$.

When $\lambda _{1}=0$, from Eqs. (\ref{real-st}, \ref{gam-1}, \ref{uniq-1}, %
\ref{rel-2}, \ref{rel-3}), we obtain%
\begin{eqnarray}
|\psi \rangle &=&(\lambda _{0},0,\lambda _{2},\lambda _{3},\lambda _{4}), \\
\gamma &=&-\lambda _{2}\lambda _{3}, \\
\iota &=&\lambda _{2}\lambda _{3}(1-1/\varrho ^{2})/\lambda _{4},
\label{rel-1} \\
\varrho &=&\sqrt{(\lambda _{0}\lambda _{4})^{2}+(\lambda _{2}\lambda
_{3})^{2}}/\sqrt{(\lambda _{2}^{2}+\lambda _{4}^{2})(\lambda
_{3}^{2}+\lambda _{4}^{2})}  \notag \\
&& \\
|\psi _{\varrho ,\iota }\rangle &=&((1/\varrho )\lambda _{0},\varrho \iota
,\varrho \lambda _{2},\varrho \lambda _{3},\varrho \lambda _{4}),
\label{equ-1}
\end{eqnarray}

\ \textit{Conclusion 2}. \ (i). For the 4-LBPS state $|\psi \rangle $ with $%
\lambda _{1}=0$, a calculation yields that$\ \iota =0$ if and only if $%
\varrho =1$ if and only if $\lambda _{0}=1/\sqrt{2}$ ( i.e. $|\psi \rangle $
is of the form $\left( 1/\sqrt{2},0,\lambda _{2},\lambda _{3},\lambda
_{4}\right) $) \cite{dli-jpa}.

(ii). The contrapositive version of the above (i) leads to the following.
For the 4-LBPS state $|\psi \rangle $ with $\lambda _{1}=0$, $\iota \neq 0$
if and only if $\varrho \neq 1$ if and only if $\lambda _{0}\neq 1/\sqrt{2}$
(i.e. $|\psi \rangle =$ $\left( \lambda _{0}(\neq 1/\sqrt{2}),0,\lambda
_{2},\lambda _{3},\lambda _{4}\right) $).

\textit{Result 2}. In light of (i) of Proposition 3 in \cite{dli-jpa}, $%
|\psi ^{\prime }\rangle $\ is LU equivalent to $|\psi \rangle $ in Eq. (\ref%
{real-st}) if and only if $|\psi ^{\prime }\rangle =|\psi _{\varrho ,\iota
}\rangle $ in Eq. (\ref{rel-3}).

From Result 2, we have the following corollary 2.1.

\textit{Corollary 2.1}. If $|\psi ^{\prime }\rangle $\ is LU equivalent to $%
|\psi \rangle $ in Eq. (\ref{real-st}), then $|\psi ^{\prime }\rangle $ is
also real because $\iota $ is real and $|\ln \varrho |=|\ln \varrho ^{\prime
}|$.

\subsubsection{Two states with different number of LBPS may be LU
equivalent\ \ \ \ }

When $\varrho \neq 1$, then $\iota $ in Eq. (\ref{rel-1}) is a non-zero real
number and $|\psi _{\varrho ,\iota }\rangle $ in Eq. (\ref{equ-1}) is a real
5-LBPS state. Thus, the 4-LBPS state $\left( \lambda _{0}(\neq 1/\sqrt{2}%
),0,\lambda _{2},\lambda _{3},\lambda _{4}\right) $ is LU equivalent to a
real 5-LBPS state $|\psi _{\varrho ,\iota }\rangle $. It means that the
number of LBPS is not a LU invariant.

For example, let
\begin{eqnarray}
|\phi \rangle &=&(1/2)(1,0,1,1,1), \\
|\phi ^{\prime }\rangle &=&(\frac{\sqrt{2}}{2},-\frac{\sqrt{2}}{4},\frac{%
\sqrt{2}}{4},\frac{\sqrt{2}}{4},\frac{\sqrt{2}}{4}).
\end{eqnarray}%
Clearly, $|\phi \rangle $ has four LBPS and $|\phi ^{\prime }\rangle $ has
five LBPS.

Moreover, a calculation yields that
\begin{equation}
|\phi ^{\prime }\rangle =H\otimes H\otimes H|\phi \rangle ,
\end{equation}%
where $H$ is the Hadmard matrix. It is well known that the Hadmard matrix is
a unitary matrix. Therefore, $|\phi ^{\prime }\rangle $ and $|\phi \rangle $
are LU equivalent though they have different number of LBPS.

\subsubsection{LU classification of real states with $\protect\gamma \protect%
\lambda _{2}\protect\lambda _{3}\neq 0$}

All the real states with $\gamma \lambda _{2}\lambda _{3}\neq 0$ \ are
partitioned into the real families $R_{1}$ and $R_{2}$. Let $R_{1}$ be the
family consisting of the 5-LBPS real states with $\gamma \neq 0$ and $\iota
\neq 0$. \ Let $R_{2}$ be the family consisting of the 4-LBPS real states of
the form $(\lambda _{0},0,\lambda _{2},\lambda _{3},\lambda _{4})$ (of
course, \ $\gamma \neq 0$) and the 5-LBPS real states with $\gamma \neq 0$
and $\iota =0$. Ref. Table II (a).

We next argue that $R_{1}$ and $R_{2}$ are LU inequivalent.

Let $|\psi \rangle $ be a state in $R_{1}$. Then, $|\psi \rangle $ is a
5-LBPS\ state with $\gamma \iota \neq 0$. In light of Result 2, if $|\psi
^{\prime }\rangle $ is LU equivalent to $|\psi \rangle $, then $|\psi
^{\prime }\rangle =|\psi _{\varrho ,\iota }\rangle $ in Eq. (\ref{rel-3}).
Thus, $|\psi ^{\prime }\rangle $ is also a real 5-LBPS\ state with $\iota
^{\prime }=\delta \varrho \lambda _{1}\neq 0$ (ref. Eq. (\ref{iota-3})).
Then, it is not hard to see that $R_{1}$ and $R_{2}$ are LU inequivalent.

Table II (a) Real families $R_{1}$ and $R_{2}$ for which $\gamma \neq 0$ and
$\lambda _{2}\lambda _{3}\neq 0$

\begin{tabular}{|l|l|}
\hline
$R_{1}$ & 5-LBPS real states with $\gamma \neq 0$ and $\iota \neq 0$ \\
\hline
$R_{2}$ & 4-LBPS real states with $\lambda _{1}=0$ \\ \hline
& and 5-LBPS real states with $\gamma \neq 0$ and $\iota =0$ \\ \hline
\end{tabular}

Family $R_{i}$ is divided into two subfamilies $R_{i}^{\prime }$ (consisting
of the states with $\varrho =1$) and $R_{i}^{\prime \prime }$ (consisting of
the states with $\varrho \neq 1$). Ref. Table II (b). Corollary 2.1 implies
that $R_{i}^{\prime }$ and $R_{i}^{\prime \prime }$ are LU\ inequivalent.
One can know that each LU class of $R_{i}^{\prime }$ is a singleton and each
LU class of $R_{i}^{\prime \prime }$ consists of only two states $|\psi
\rangle $ and $|\psi _{\varrho ,\iota }\rangle $.

Table II (b). Family $R_{i}$ is divided into two subfamilies $R_{i}^{\prime
} $ (the states with $\varrho =1$) and $R_{i}^{\prime \prime }$ (the states
with $\varrho \neq 1$).

\begin{tabular}{|l|l|}
\hline
$\varrho =1$ & $\varrho \neq 1$, \\ \hline
$R_{1}^{\prime }=$\{$(\lambda _{0},\delta \lambda _{1},\lambda _{2},\lambda
_{3},\lambda _{4})$\} & $R_{1}^{\prime \prime }=$\{$|\psi \rangle $, $|\psi
_{\varrho ,\iota }\rangle $\} \\ \hline
$R_{2}^{\prime }=$\{$(\frac{1}{\sqrt{2}},0,\lambda _{2},\lambda _{3},\lambda
_{4})$\} & $R_{2}^{\prime \prime }=$ \{$|\psi \rangle $, $|\psi _{\varrho
,\iota }\rangle $\} \\ \hline
\end{tabular}

In light of Result 2 and Corollary 2.1, we have the following Corollary 2.2.

\textit{Corollary 2.2.} ASD of a real state with $\gamma \lambda _{2}\lambda
_{3}\neq 0$ is unique if and only if $\varrho =1$. In other words, via Eq. (%
\ref{asd-unique-1}), ASD of a real state with $\gamma \lambda _{2}\lambda
_{3}\neq 0$ is unique if and only if $\lambda _{0}^{2}+\lambda _{1}^{2}=%
\frac{1}{2}+\frac{\delta \lambda _{1}\lambda _{2}\lambda _{3}}{\lambda _{4}}$%
.

The contrapositive version of Corollary 2.2 leads to the following. ASD of a
real state with $\gamma \lambda _{2}\lambda _{3}\neq 0$ is not unique if and
only if $\varrho \neq 1$. In other words, ASD of a real state with $\gamma
\lambda _{2}\lambda _{3}\neq 0$ is not unique if and only if $\lambda
_{0}^{2}+\lambda _{1}^{2}\neq \frac{1}{2}+\frac{\delta \lambda _{1}\lambda
_{2}\lambda _{3}}{\lambda _{4}}$.

\subsubsection{The number of LBPS\ is not LU invariant for $R_{2}^{\prime
\prime }$}

In light of (ii) of Conclusion 1, the 5-LBPS real states with $\gamma \neq 0$
and $\iota =0$ belong to $R_{2}^{\prime \prime }$.

In light of (i) of Conclusion 2, $R_{2}^{\prime }$ consists of the states of
only the form $(\frac{1}{\sqrt{2}},0,\lambda _{2},\lambda _{3},\lambda _{4})$%
.

In light of (ii) of Conclusion 2, the states\ of the form $(\lambda
_{0}(\neq 1/\sqrt{2}),0,\lambda _{2},\lambda _{3},\lambda _{4})$ belong to $%
R_{2}^{\prime \prime }$.

In light of Conclusions 1 and 2, each LU class of $R_{2}^{\prime \prime }$
is a pair of a 4-LBPS state $|\psi \rangle $\ of the form $(\lambda
_{0}(\neq 1/\sqrt{2}),0,\lambda _{2},\lambda _{3},\lambda _{4})$ and a real
5-LBPS state with $\gamma \neq 0$ and $\iota =0$ ($=$ $|\psi _{\varrho
,\iota }\rangle $). For example, ($|\phi \rangle $, $|\phi ^{\prime }\rangle
$) is a LU class of $R_{2}^{\prime \prime }$.

Therefore, the number of LBPS is not LU invariant for $R_{2}^{\prime \prime
} $.

\subsection{LU classification of complex states with $\protect\gamma \neq 0$
and $\protect\lambda _{2}\protect\lambda _{3}=0$}

Let $|\psi \rangle $ be the state with $\gamma \neq 0$ and $\lambda
_{2}\lambda _{3}=0$. Then, clearly $\lambda _{1}\neq 0$ for the states with $%
\gamma \neq 0$ and $\lambda _{2}\lambda _{3}=0$.

\subsubsection{Calculating $\protect\varrho $, $\protect\iota $, and $|%
\protect\psi _{\protect\varrho ,\protect\iota }\rangle $}

From Appendix A, when $\lambda _{1}\neq 0$ and $\lambda _{2}\lambda _{3}=0$,
the following two states are LU equivalent for any $\varphi $ and $\chi $.
\begin{eqnarray}
|\psi \rangle &=&(\lambda _{0},\lambda _{1}e^{i\varphi },\lambda
_{2},\lambda _{3},\lambda _{4}), \\
|\varpi \rangle &=&(\lambda _{0},\lambda _{1}e^{i\chi },\lambda _{2},\lambda
_{3},\lambda _{4}),
\end{eqnarray}%
\ Note that $\varrho (|\psi \rangle )=\varrho (|\varpi \rangle )$. It
implies that a state with $\gamma \neq 0$ and $\lambda _{2}\lambda _{3}=0$
has infinite ASD. Appendix A tells us that we don't need to consider the
phases when determining if two states with $\gamma \neq 0$ and $\lambda
_{2}\lambda _{3}=0$ are LU \ equivalent. That is, we only need to consider
the following states with $\gamma \neq 0$ and $\lambda _{2}\lambda _{3}=0$.
\begin{equation}
|\psi \rangle =(\lambda _{0},\lambda _{1},\lambda _{2},\lambda _{3},\lambda
_{4})  \label{p-2}
\end{equation}

For $|\psi \rangle $ in Eq. (\ref{p-2}) with $\gamma \neq 0$ and $\lambda
_{2}\lambda _{3}=0$, a calculation yields that
\begin{eqnarray}
\iota &=&\lambda _{1}/\varrho ^{2},  \label{pp-1} \\
\varrho &=&\sqrt{\lambda _{0}^{2}+\lambda _{1}^{2}}/\sqrt{1-\lambda
_{0}^{2}-\lambda _{1}^{2}},  \label{pp-2} \\
|\psi _{\varrho ,\iota }\rangle &=&((1/\varrho )\lambda _{0},\lambda
_{1}/\varrho ,\varrho \lambda _{2},\varrho \lambda _{3},\varrho \lambda
_{4}).  \label{p-3}
\end{eqnarray}

When $\varrho =1$, from Eqs. (\ref{pp-1}, \ref{pp-2}, \ref{p-3}), obtain
\begin{eqnarray}
\iota &=&\lambda _{1}, \\
\lambda _{0}^{2}+\lambda _{1}^{2} &=&1/2,  \label{p-4} \\
|\psi _{\varrho ,\iota }\rangle &=&|\psi \rangle .
\end{eqnarray}

\textit{Result 3}. In light of (ii) of Proposition 3 in \cite{dli-jpa}, one
can know that $|\psi ^{\prime }\rangle $ is LU equivalent to $|\psi \rangle $
with $\gamma \neq 0$ and $\lambda _{2}\lambda _{3}=0$ if and only if $|\psi
^{\prime }\rangle =|\psi _{\varrho ,\iota }\rangle $ ignoring the phases.

From Result 3, we have the following corollary 3.1.

\textit{Corollary 3.1}. If $|\psi ^{\prime }\rangle $ is LU equivalent to $%
|\psi \rangle $ with $\gamma \neq 0$ and $\lambda _{2}\lambda _{3}=0$, then $%
|\ln \varrho |=|\ln \varrho ^{\prime }|$ and$\ \ |\psi ^{\prime }\rangle $
and $|\psi \rangle $ have the same kinds of LBPS.

\subsubsection{LU classification of complex states with $\protect\gamma \neq
0$ and $\protect\lambda _{2}\protect\lambda _{3}=0$}

In Table III (a), we partition the complex states with$\ \gamma \neq 0$, and
$\lambda _{2}\lambda _{3}=0$\ into three families $C_{i}$, $i=1,2,3$.

We next argue that $C_{i}$, $i=1,2,3$, are LU inequivalent.

In light of Result 3 and via Eqs. (\ref{p-2}, \ref{p-3}), one can know that $%
\lambda _{i}$ and $\varrho \lambda _{i}$, $i=2,3$, both vanish or neither
does. But, $\lambda _{2}=0$ for $C_{1}$ and $C_{3}$\ while $\lambda _{2}\neq
0$ for $C_{2}$, and $\lambda _{3}=0$\ \ for $C_{2}$ and $C_{3}$\ while $%
\lambda _{3}\neq 0$ for $C_{1}$. Therefore, $C_{i}$, $i=1,2,3$, are LU
inequivalent.

Each complex Family $C_{i}$ is divided into two subfamilies $C_{i}^{\prime }$
(states with $\varrho =1$) and $C_{i}^{\prime \prime }$ (states with $%
\varrho \neq 1$). Note that each LU class includes infinite states with $%
\gamma \neq 0$ and $\lambda _{2}\lambda _{3}=0$. After ignoring phase, each
LU class of $C_{i}^{\prime }$ is a singleton and each LU class of $%
C_{i}^{\prime \prime }$ consists of only two states $|\psi \rangle $ and $%
|\psi _{\varrho ,\iota }\rangle $. Ref. Table III (a). Corollary 3.1 implies
that $C_{i}^{\prime }$ and $C_{i}^{\prime \prime }$, $i=1,2,3$, are LU
inequivalent.

In light of Result 3 and Corollary 3.1, we have the following Corollary 3.2.

\textit{Corollary 3.2}. Ignoring phases, ASD of a complex state with $\gamma
\neq 0$ and $\lambda _{2}\lambda _{3}=0$ is unique if and only if $\varrho
=1 $. In other words, via Eq. (\ref{p-4}), ASD of a complex state with $%
\gamma \neq 0$ and $\lambda _{2}\lambda _{3}=0$ is unique if and only if $%
\lambda _{0}^{2}+\lambda _{1}^{2}=1/2$ ignoring phases.

The contrapositive version of Corollary 3.2 leads to the following. ASD of a
complex state with $\gamma \neq 0$ and $\lambda _{2}\lambda _{3}=0$ is not
unique if and only if $\ \varrho \neq 1\ $(in other words,$\ \lambda
_{0}^{2}+\lambda _{1}^{2}\neq 1/2$) ignoring phases.

Table III (a). Complex families\textbf{\ }$C_{1},$\textbf{\ }$C_{2},$and $%
C_{3}$ for which $\gamma \neq 0$ and $\lambda _{2}\lambda _{3}=0$

\begin{tabular}{|l|l|l|}
\hline
$\gamma \neq 0,\lambda _{2}\lambda _{3}=0$ & $\varrho =1$ & $\varrho \neq 1$
\\ \hline
$C_{1}$;\{$(\lambda _{0},\lambda _{1}e^{i\varphi },0,\lambda _{3},\lambda
_{4})$\} & $C_{1}^{\prime }$;$\vartriangleleft $ & $C_{1}^{\prime \prime }$%
;\{$|\psi \rangle $, $|\psi _{\varrho ,\iota }\rangle $\} \\ \hline
$C_{2}$;\{$(\lambda _{0},\lambda _{1}e^{i\varphi },\lambda _{2},0,\lambda
_{4})$\} & $C_{2}^{\prime }$;$\vartriangleleft $ & $C_{2}^{\prime \prime }$%
;\{$|\psi \rangle $, $|\psi _{\varrho ,\iota }\rangle $\} \\ \hline
$C_{3}$;\{$(\lambda _{0},\lambda _{1}e^{i\varphi },0,0,\lambda _{4})$\} & $%
C_{2}^{\prime }$;$\vartriangleleft $ & $C_{2}^{\prime \prime }$;\{$|\psi
\rangle $, $|\psi _{\varrho ,\iota }\rangle $\} \\ \hline
\end{tabular}

$\vartriangleleft $ $\lambda _{0}^{2}+\lambda _{1}^{2}=1/2$.

\subsection{LU classification of complex 5-LBPS states with the phases $%
\protect\varphi \neq 0$ or $\protect\pi $}

Let $|\psi \rangle $ be a complex 5-LBPS state with the phases $\varphi \neq
0$ or $\pi $.\ We write $|\psi \rangle $ as follows.

\begin{equation}
|\psi \rangle =(\lambda _{0},\lambda _{1}e^{i\varphi },\lambda _{2},\lambda
_{3},\lambda _{4}),  \label{c4-1}
\end{equation}

\subsubsection{Calculating $\protect\varrho $, $\protect\iota $, and the
state $|\protect\psi _{\protect\varrho ,\protect\iota }\rangle $}

For the complex 5-LBPS states with the phases $\varphi \neq 0$ or $\pi $,
via Eqs. (\ref{gama-1}, \ref{fact}, \ref{factor}, \ref{eq-state-1}), we have
the following
\begin{eqnarray}
\gamma &=&\lambda _{1}\lambda _{4}e^{i\varphi }-\lambda _{2}\lambda _{3}, \\
\iota &=&(\lambda _{2}\lambda _{3}+\gamma ^{\ast }/\varrho ^{2})/\lambda
_{4}, \\
\varrho &=&\sqrt{J_{4}+J_{1}}/\sqrt{(\lambda _{2}^{2}+\lambda
_{4}^{2})(\lambda _{3}^{2}+\lambda _{4}^{2})}, \\
|\psi _{\varrho ,\iota }\rangle &=&((1/\varrho )\lambda _{0},\varrho \iota
,\varrho \lambda _{2},\varrho \lambda _{3},\varrho \lambda _{4}).
\label{c4-2}
\end{eqnarray}

It is not hard to see that $\gamma \neq 0$ and the imaginary part of $\iota $
does not vanish. Thus, $|\psi _{\varrho ,\iota }\rangle $ is also a complex
5-LBPS state whose phase is not $0$ or $\pi $.

When $\varrho =1$, from the above equations a calculation yields that
\begin{eqnarray}
\iota &=&\lambda _{1}e^{-i\varphi }, \\
|\psi _{\varrho ,\iota }\rangle &=&(\lambda _{0},\lambda _{1}e^{-i\varphi
},\lambda _{2},\lambda _{3},\lambda _{4}) \\
&=&|\psi ^{\ast }\rangle  \label{unique-4} \\
&\neq &|\psi \rangle .
\end{eqnarray}%
where $|\psi ^{\ast }\rangle $ is the complex conjugate of $|\psi \rangle $.

\textit{Result 4}.

\ In light of (i) of Proposition 3 in \cite{dli-jpa}, $|\psi ^{\prime
}\rangle $ ($\neq $ $|\psi \rangle $)\ is LU equivalent to the complex
5-LBPS state $|\psi \rangle $ with the phase $\varphi \neq 0$ or $\pi $\ if
and only if $|\psi ^{\prime }\rangle =|\psi _{\varrho ,\iota }\rangle $ in
Eq. (\ref{c4-2}).

From Result 4, we have the following Corollaries 4.1.

\ \textit{Corollary 4.1}. If $|\psi ^{\prime }\rangle $\ is LU equivalent to
the complex 5-LBPS state $|\psi \rangle $ with the phase $\varphi \neq 0$ or
$\pi $, then $|\ln \varrho |=|\ln \varrho ^{\prime }|$ and $|\psi ^{\prime
}\rangle $ is also a complex 5-LBPS state whose phase is not $0$ or $\pi $.

\subsubsection{LU classification of complex 5-LBPS states with the phases $%
\protect\varphi \neq 0$ or $\protect\pi $}

Let Family $C_{4}$\ consist of complex 5-LBPS\ states with the phases $%
\varphi \neq 0$ or $\pi $. It implies that for those states, $\gamma \neq 0$
and $\iota \neq 0$.

We next argue that $C_{4}$ is LU inequivalent to $C_{i}$, $i=1,2,3$.

In\ light of Corollary 4.1, if $|\psi ^{\prime }\rangle $\ is LU equivalent
to the complex 5-LBPS state $|\psi \rangle $ with the phase $\varphi \neq 0$
or $\pi $, then $|\psi ^{\prime }\rangle $ is also a complex 5-LBPS state.
From Table III (a), one can see that $C_{i}$, $i=1,2$, consist of 4-LBPS
states and $C_{3}$ consists of 3-LBPS\ states. Therefore, $C_{4}$ is LU
inequivalent to $C_{i}$, $i=1,2,3$.

$C_{4}$ is divided into two subfamilies $C_{4}^{\prime }$ (states with $%
\varrho =1$) and $C_{4}^{\prime \prime }$ (states with $\varrho \neq 1$).
Ref. Table III (b). Via Eq. (\ref{unique-4}) and in light of Result 4, each
LU class of $C_{4}^{\prime }$ consists of a state and its complex conjugate,
while each LU class of $C_{4}^{\prime \prime }$ consists of only two states $%
|\psi \rangle $ and $|\psi _{\varrho ,\iota }\rangle $, where $|\psi
_{\varrho ,\iota }\rangle \neq |\psi ^{\ast }\rangle $.

In light of Result 4 and Corollary 4.1, we have the following Corollary 4.2.

\textit{Corollary 4.2}. Considering $|\psi \rangle $ and its complex
conjugate $|\psi ^{\ast }\rangle $ to be the same, ASD of a complex 5-LBPS
state with the phases $\varphi \neq 0$ or $\pi $ is unique if and only if $%
\varrho =1$.

The contrapositive version of Corollary 4.2 leads to the following.
Considering $|\psi \rangle $ and its complex conjugate $|\psi ^{\ast
}\rangle $ to be the same, ASD of a complex 5-LBPS state with the phases $%
\varphi \neq 0$ or $\pi $ is not unique if and only if $\varrho \neq 1$.

Table III (b). Complex Family $C_{4}$ (5-LBPS states with $\varphi \neq 0$
or $\pi $).

\begin{tabular}{|l|l|}
\hline
$C_{4}$: 5-LBPS\ states with $\varphi \neq 0$,$\pi $ & each LU class \\
\hline
$C_{4}^{\prime }$ $=$\{ states with $\varrho =1$\} & $=$\{$|\psi \rangle
,|\psi ^{\ast }\rangle $\} \\ \hline
$C_{4}^{\prime \prime }=$\{states with $\varrho \neq 1$\} & $=$\{$|\psi
\rangle ,|\psi _{\varrho ,\iota }\rangle $\} \\ \hline
\end{tabular}

\subsubsection{$C_{4}^{\prime }$ ($C_{4}^{\prime \prime }$) is the set of
states being (not being) LU equivalent to their complex conjugates}

For the 5-LBPS state $|\psi \rangle $ with $\varrho (|\psi \rangle )\neq 1$,
one can also verify that $|\psi _{\varrho ,\iota }\rangle $ is not $|\psi
^{\ast }\rangle $ as follows. Clearly, $\varrho (|\psi \rangle )=\varrho
(|\psi ^{\ast }\rangle )$, thus $\varrho (|\psi \rangle )\varrho (|\psi
^{\ast }\rangle )\neq 1$ when $\varrho (|\psi \rangle )\neq 1$. Via Eq. (\ref%
{invar}), $|\psi _{\varrho ,\iota }\rangle $ is not $|\psi ^{\ast }\rangle $%
. Thus, each state of $C_{4}^{\prime \prime }$ is LU inequivalent to its
complex conjugate.

A twelfth degree complex polynomial invariant $I_{6}$, introduced by Grassl
\cite{Grassl}, can distinguish among two complex conjugate ASD states which
are LU inequivalent \cite{Acin01}. The bipartite operational measure $E_{1}$
is also used to determine if a state is LU equivalent to its complex
conjugate \cite{Vicente}. For a complex 5-LBPS state $|\psi \rangle $ with
the phases $\varphi \neq 0$ or $\pi $, the value of $\varrho (|\psi \rangle
) $ determines whether or not $|\psi \rangle $ is LU equivalent to its
complex conjugate $|\psi ^{\ast }\rangle $. Note that $\varrho $ is positive
and simpler than $I_{6}$ and $E_{1}$.

It is known that $|\psi \rangle $ and $|\psi ^{\ast }\rangle $ possess the
same entanglement properties \cite{Kraus-prl}.\ The class of states not
being LU equivalent to their complex conjugates is referred to as NCLU \cite%
{Vicente}. The existence of NCLU states is a surprising property of
multipartite system which does not exist for the bipartite system \cite%
{Vicente}. Clearly, one can see that $C_{4}^{\prime \prime }$ is just NCLU,
and it is easy to find $C_{4}^{\prime \prime }$.

\textit{Remark 1:} 5-LBPS states are partitioned into four families : the
positive family $P_{1}$, the real families $R_{1}$ and $R_{2}$, and the
complex family $C_{4}$. Furthermore, each family is divided into two
subfamilies. Thus, 5-LBPS states are partitioned into seven subfamilies: $%
P_{1}^{\prime }$, $P_{1}^{\prime \prime }$, $R_{1}^{\prime }$, $%
R_{1}^{\prime \prime }$, $R_{2}^{\prime \prime }$, $C_{4}^{\prime }$ and $%
C_{4}^{\prime \prime }$.

\textit{Remark 2.} The number of LBPS is a LU invariant for the GHZ SLOCC\
class except for only $R_{2}^{\prime \prime }$. Thus, two states of the GHZ
SLOCC\ class except for $R_{2}^{\prime \prime }$ with different number of
LBPS are LU inequivalent.

\textit{Remark 3}. For any state of the GHZ SLOCC\ class, when $\varrho \neq
1$ then its ASD is not unique, while $\varrho =1$, its ASD is unique for the
families $P_{i},i=1,2,3,4$, $R_{1}$, and $R_{2}$, for the families $C_{i}$, $%
i=1,2,3$, ignoring the phases, and for the family $C_{4}$ considering the
state and its complex conjugate to be the same. Therefore, for the GHZ
SLOCC\ class, ASD\ is unique if and only if $\varrho =1$. Thus, we can call $%
\varrho $ the uniqueness parameter. When $\varrho \neq 1$, from $|\psi
\rangle $\ and \ $|\psi _{\varrho ,\iota }\rangle $, we choose the one with $%
\varrho <1$ as the canonical ASD.

\subsection{The argument for the complete LU classification of the GHZ
SLOCC\ class}

We partition the positive states with $\gamma =0$ into four positive
families $P_{i},i=1,2,3,4$, the real states with $\gamma \neq 0$ and $%
\lambda _{2}\lambda _{3}\neq 0$ into two real families $R_{1}$ and $R_{2}$,
the complex states with $\gamma \neq 0$ and $\lambda _{2}\lambda _{3}=0$
into three complex families $C_{i}$, $i=1,2,3$,$\ $and let $C_{4}$ include
the complex 5-LBPS states with $\varphi \neq 0$ or $\pi $. Note that for
5-LBPS states with $\varphi \neq 0$ or $\pi $, $\gamma \neq 0$ and $\iota
\neq 0$.

In total, we partition the GHZ\ SLOCC\ class of three qubits into 10
families. Each family is partitioned into two subfamilies one of which has $%
\varrho =1$ while the other one has $\varrho \neq 1$.

(i). Since $J_{1}$ is LU invariant, where $J_{1}=|\gamma |^{2}$, and $\gamma
=0$ for $P_{1},P_{2},P_{3}$, and $P_{4}$ while $\gamma \neq 0$ for $R_{1}$, $%
R_{2}$, $C_{1}$, $C_{2}$, $C_{3}$, and $C_{4}$, the positive families $%
P_{1},P_{2},P_{3}$, and $P_{4}$ are LU inequivalent to $R_{1}$, $R_{2}$, $%
C_{1}$, $C_{2}$, $C_{3}$, and $C_{4}$.

(ii). In light of Results 2 and 3, the real families $R_{1}$ and $R_{2}$ are
LU inequivalent to the complex families $C_{i}$, $i=1,2,3$. For any state in
$R_{1}$ and $R_{2}$, the phase is $0$ or $\pi $, while for any state of $%
C_{4}$, the phase is neither $0$ nor $\pi $. In light of Results 3 and 4, $%
C_{4}$ is LU inequivalent to the real families $R_{1}$ and $R_{2}$.

\subsection{A LU invariant for the GHZ\ SLOCC\ class}

In the above section, for the positive states (resp. the real states and the
complex states) of the GHZ SLOCC\ class, we show that $|\ln \varrho |$ is a
LU invariant. Thus, the state with $\varrho =1$ and the state with $\varrho
\neq 1$ are LU inequivalent. Then, we can conclude that

(1). $|\ln \varrho |$ is a LU invariant for the whole GHZ\ SLOCC\ class.

(2). For any two states $|\psi _{1}\rangle $ and $|\psi _{2}\rangle $ of the
GHZ SLOCC\ class, if $\varrho (|\psi _{1}\rangle )\varrho (|\psi _{2}\rangle
)\neq 1$, then $|\psi _{1}\rangle $ and $|\psi _{2}\rangle $ are LU
inequivalent.

We propose $\frac{1}{1+|\ln \varrho |}$ as a measure of the entanglement for
the GHZ\ SLOCC\ class. For the measure, the GHZ state has the maximal
entanglement $\frac{1}{1+|\ln \varrho |}=1$. For $|\phi \rangle $, $\varrho
=1/\sqrt{2}$ and for $|\phi ^{\prime }\rangle $, $\varrho ^{\prime }=\sqrt{2}
$. Thus, for $|\phi \rangle $ and $|\phi ^{\prime }\rangle $, $\frac{1}{%
1+|\ln \varrho |}=\frac{1}{1+\sqrt{2}}$.

\subsection{Some states with the unique ASD}

For the positive or real state $|\psi \rangle $ with $\varrho =1$, subjected
to local random unitary noise, the ASD of $|\psi \rangle $ does not change.
That is, $U_{1}\otimes U_{2}\otimes U_{3}|\psi \rangle $ and $|\psi \rangle $
have the same ASD.

We give the following positive states for which $\varrho =1$.

\begin{eqnarray*}
&&(1/\sqrt{2})(|000\rangle +|111\rangle ) \\
&&(1/\sqrt{2})|000\rangle +(1/2)|101\rangle +(1/2)|111\rangle \\
&&(1/\sqrt{2})|000\rangle +(1/2)|110\rangle +(1/2)|111\rangle \\
&&\frac{1}{\sqrt{2}}|000\rangle +\frac{1}{2\sqrt{2}}|101\rangle +\frac{1}{2%
\sqrt{2}}|110\rangle +\frac{1}{2}|111\rangle \\
&&\frac{1}{\sqrt{2}}|000\rangle +\frac{1}{2\sqrt{2}}(|100\rangle
+|101\rangle +|110\rangle +|111\rangle )
\end{eqnarray*}

\section{Summary}

It is well known that pure states of three qubits are partitioned into six
SLOCC\ equivalence classes, two of which are the W SLOCC equivalence class
and the GHZ SLOCC equivalence\ class \cite{Dur}. Ac\'{\i}n et al.
partitioned pure states of three qubits into five types \cite{Acin00}. The
positive states of the GHZ SLOCC\ class were partitioned into four
subclasses \cite{Kumari}.

We propose the LU invariant $|\ln \varrho |$ and the entanglement measure $%
\frac{1}{1+|\ln \varrho |}$ for the GHZ\ SLOCC equivalence\ class of three
qubits. Via parameters $\varrho $, $\iota $, and $\gamma $, we partition
positive, real, and complex states of the GHZ SLOCC class into ten families,
and each family into two subfamilies under LU.

Each LU class of $C_{4}^{\prime }$ is a pair of a complex 5-LBPS state with
the phases $\varphi \neq 0$ or $\pi $\ and its complex conjugate. $%
C_{4}^{\prime \prime }$ is the set of states not being LU equivalent to
their complex conjugates. But, $C_{4}^{\prime \prime }$ does not exist for
the bipartite system. It is interesting to find criteria to partition $%
C_{4}^{\prime }$ and $C_{4}^{\prime \prime }$ furthermore.

Each LU class of $R_{2}^{\prime \prime }$ is a pair of a 4-LBPS state $|\psi
\rangle $\ of the form $(\lambda _{0}(\neq 1/\sqrt{2}),0,\lambda
_{2},\lambda _{3},\lambda _{4})$ and a real 5-LBPS state with $\gamma \neq 0$
and $\iota =0$ ($=$ $|\psi _{\varrho ,\iota }\rangle $). We show that the
number of LBPS is a LU invariant for the GHZ SLOCC\ class except for only $%
R_{2}^{\prime \prime }$.

We show that for the GHZ SLOCC\ class, ASD of a state is unique if and only
if $\varrho =1$. Thus, subjected to local random unitary noise, ASD of a
state with $\varrho =1$ does not change. We give some positive states with $%
\varrho =1$.

A. Kumari and S. Adhikari partitioned positive states (i.e. the states with
the phase factor $\theta =0$) of the GHZ SLOCC\ class into four subclasses.
Via this LU classification of the GHZ SLOCC\ class, it is easy to see that
the four subclasses are inequivalent under LU.

\section{Appendix A. LU equivalence of some special ASD states}


\setcounter{equation}{0} \renewcommand{\theequation}{B\arabic{equation}}

Let
\begin{equation}
|\psi \rangle =(\lambda _{0},\lambda _{1}e^{i\omega },\lambda _{2},\lambda
_{3},\lambda _{4}),
\end{equation}

\begin{equation}
|\psi ^{\prime }\rangle =(\lambda _{0},\lambda _{1}e^{i\omega ^{\prime
}},\lambda _{2},\lambda _{3},\lambda _{4}).
\end{equation}

When $\lambda _{2}\lambda _{3}=0$ and $\lambda _{0}\lambda _{1}\lambda
_{4}\neq 0$, we can show that $|\psi ^{\prime }\rangle $ is LU equivalent to
$|\psi \rangle $. \

Case 1. $\lambda _{3}=0$ and $\lambda _{0}\lambda _{1}\lambda _{2}\lambda
_{4}\neq 0$ \ Let
\begin{eqnarray}
U^{A} &=&diag\left( e^{i\phi _{1}},e^{i(2\phi _{1}+\phi _{2})}\right) ,
\notag \\
U^{B} &=&diag\left( e^{-i\phi _{1}},e^{-i\phi _{1}}\right) ,  \notag \\
U^{C} &=&diag(1,e^{i(-\phi _{2}-\phi _{1})}).  \label{U-ABC}
\end{eqnarray}%
where $\phi _{2}+\phi _{1}=\omega ^{\prime }-\omega $. Via $U^{A}$, $U^{B}$,
and $U^{C}$ in Eq. (\ref{U-ABC}), a calculation yields that $|\psi ^{\prime
}\rangle =U^{A}\otimes U^{B}\otimes U^{C}|\psi \rangle $. Therefore, $|\psi
^{\prime }\rangle $ is LU equivalent to $|\psi \rangle $.

Case 2. $\lambda _{2}=0$ and $\lambda _{0}\lambda _{1}\lambda _{3}\lambda
_{4}\neq 0$ or $\lambda _{2}=\lambda _{3}=0$ and $\lambda _{0}\lambda
_{1}\lambda _{4}\neq 0$. Let
\begin{eqnarray*}
U^{A} &=&diag\left( e^{i\alpha },e^{i\beta }\right) , \\
U^{B} &=&diag\left( e^{-i\alpha },e^{-i\beta }\right) , \\
U^{C} &=&I, \\
\beta -\alpha &=&\omega ^{\prime }-\omega .
\end{eqnarray*}

Then, $|\psi ^{\prime }\rangle =U^{A}\otimes U^{B}\otimes U^{C}|\psi \rangle
$.

\end{document}